\documentclass[runningheads,a4paper,english,fleqn,orivec]{llncs}%
\usepackage{eso-pic}
\AddToShipoutPicture*{
\put(133,160){
\scriptsize
Proceedings of WLPE 2013, {\tt arXiv:1308.2055}, August 2013.
}}
\newcommand{\submitted}[1]{}
\newcommand{\revised}[1]{}
\newcommand{\accepted}[1]{}

\usepackage[nodate]{datetime}
\usepackage{babel}
\usepackage{amsmath}
\usepackage{mathtools}[2011/02/12]
\usepackage{amsfonts,amssymb,stmaryrd}
\usepackage{MnSymbol}
\usepackage{mdwlist}
\usepackage[applemac]{inputenc}

\usepackage{galois}

\usepackage{tikz}%\usetikzlibrary{snakes,calc,matrix,arrows}

\usepackage[scr=rsfs]{mathalfa}
\newcommand{\mathLTL}{\mathcal}
\AtBeginDocument{\renewcommand{\mathC}{\mathcal}}

\usepackage[linkcolor=blue, urlcolor=blue, backref=false, final]{hyperref}
\usepackage[fancyproofs,noextended,squareitemtag]{theorems}[2013/01/24]%,fancyexamples,final

\title{Towards an Effective Decision Procedure for LTL formulas with Constraints\thanks{This work has been partially supported by the \textsc{eu (feder)}
and the Spanish \textsc{mec/micinn}, ref.  \textsc{tin 2010-21062-c02-02},
and by Generalitat Valenciana, ref.  \textsc{prometeo2011/052}.}}
\titlerunning{Towards a Decision Method for LTL formulas with Constraints}
\author{Marco Comini\inst{1}%
\and Laura Titolo\inst{1}\and Alicia Villanueva\inst{2} }
\authorrunning{M. Comini \and L. Titolo \and A. Villanueva}

\institute{DIMI, Universit\`a degli Studi di Udine,\\
\email{\{marco.comini,laura.titolo\}@uniud.it}
\and
DSIC, Universitat Polit\`ecnica de Val\`encia\\
\email{villanue@dsic.upv.es}}

\input{wgmacros-TCCP}

\renewcommand*{\sequent}[4][]{\gdef\and{\quad\hfill}
    \gdef\ands{\;\hfill\ldots\hfill\;}\global\setlabel{#1}
    {#1}\frac{\:\:#2\:\:}{\;#3\;}\;\text{\small#4}}

\input{CompGIBUFixBehaTCCP-macros}
\newcommand{\asat}[2]{#1 \models#2}%\mathrel{\hat{}} _{\forall}
\newcommand{\notasat}[2]{#1 \nmodels #2}%\mathrel{\hat{}}_{\forall}

\newcommand*{\Lneg}[1]{\ifempty{#1}{\mathop{\dot{\neg}}}{\mathop{\dot{\neg}} #1}}
\newcommand*{\Lnext}[1]{\ifempty{#1}{\mathop{\bigcirc}}{\mathop{\bigcirc} #1}}

\newcommand*{\Luntil}[2]{\ifempty{#1}{\mathcal{U}}{#1 \mathrel{\mathcal{U}} #2}}
\newcommand*{\Lwuntil}[2]{\ifempty{#1}{\mathcal{W}}{#1 \mathrel{\mathcal{W}} #2}}
\monobioperator{\Ldisj}{\mathrel{\dot\vee}}{\mathrel{\dot\bigvee}}
\monobioperator{\Lconj}{\mathrel{\dot\wedge}}{\mathrel{\dot\bigwedge}}
\newcommand*{\Lhid}[2]{\mathop{\dot{\exists}}\nolimits_{#1} \ifempty{#2}{}{#2}}
\newcommand*{\Lalways}[1]{\ifempty{#1}{\mathop{\Box}}{\mathop{\Box} #1}}
\newcommand*{\Leventually}[1]{\ifempty{#1}{\mathop{\Diamond}}{\mathop{\Diamond} #1}}

\newcommand*{\Limpl}[2]{#1 \mathrel{\dot\rightarrow} #2}%{\ifempty{#1}{\dot\rightarrow}}
\newcommand*{\Ltrue}{\dot{\mathit{true}}}
\newcommand*{\Lfalse}{\dot{\mathit{false}}}

\newcommand*{\csltl}[1][]{\ensuremath{\textsf{csLTL}_{#1}}} %csLTL formulas
\newcommand*{\Fdom}{\mathbb{F}}
\newcommand*{\leqF}{\Limpl{}{}}

\newcommand*{\lubF}{\Ldisj}

\newcommand*{\FtoM}{\parensmathoper{\gamma^{\Fdom}}}

\newcommand*{\FAa}[3][]{\mathop{\mathLTL{A}_{\mathit{#1}}}\BBrackets[{#3}]{#2}}
\newcommand*{\FDd}[1][]{\syntaxoper{\mathLTL{D}_{\mathit{#1}}}}
 
\newcommand*{\FI}{\mathLTL{I}}
\newcommand*{\interpF}{\interpA[\Fdom]}
\newcommand*{\SF}{\mathLTL{S}}

\renewcommand*{\CShid}[2]{\mathop{\tilde{\exists}}\nolimits_{#1} #2}
\renewcommand*{\seqHid}[2]{\mathop{\tilde{\exists}}\nolimits_{#1} #2}

\newcommand*{\openleaf}{\odot}
\newcommand*{\closedleaf}{\times}
\newcommand*{\pltl}{\textsf{PLTL}}
\newcommand*{\ltl}{\textsf{LTL}}
\newcommand*{\Tname}[1][\Phi]{\mathcal{T}_{#1}}
\newcommand*{\Tlabel}{\parensmathoper{\mathit{L}}}
\newcommand*{\Tnodes}{\mathit{Nodes}}
\newcommand*{\Tinit}[1][\Phi]{n_{#1}}
\newcommand*{\Tbranches}{\mathit{B}}
\newcommand*{\Tsucc}{\mathrel{\mathit{R}}}
\newcommand*{\Ttabl}[1][\Phi]{(\Tnodes, \Tinit[#1], \Tlabel{} , \Tbranches, \Tsucc{}{} )}
\newcommand*{\Tpath}{\parensmathoper{\mathsf{path}}}
\newcommand*{\Tstages}{\parensmathoper{\mathsf{stages}}}
\newcommand*{\Tstores}{\parensmathoper{\mathsf{stores}}}
\newcommand*{\nextop}{\parensmathoper{\mathsf{next}}}

\newcommand*{\cntx}[1]{{#1}^{*}}
\newcommand*{\absat}{saturated}
\newcommand*{\clo}{\parensmathoper{\mathit{clo}}}
\newcommand*{\preclo}{\parensmathoper{\mathit{preclo}}}
\newcommand*{\subf}{\parensmathoper{\mathit{subf}}}
\newcommand*{\negctx}{\parensmathoper{\mathit{negctx}}}
\newcommand*{\conjclo}{\parensmathoper{\mathit{clo}}}
\newcommand*{\Arule}{\parensmathoper{\mathit{A}}}
\newcommand*{\Brulel}{\parensmathoper{\mathit{B}_1}}
\newcommand*{\Bruler}{\parensmathoper{\mathit{B}_2}}
\newcommand*{\Tstream}{\parensmathoper{\sigma}}
\newcommand*{\streamDep}{\parensmathoper{\mathit{dep}}}
\newcommand*{\streamHead}[2]{\ifempty{#1}{\mathit{name}}{\mathit{name}(#1,#2)}}
\newcommand*{\streq}{\mathrel{\dot{=}}}

\begin{document}

\maketitle

\begin{abstract}
This paper presents an ongoing work that is part of a more wide-ranging
project whose final scope is to define a method to validate \ltl{} formulas
\wrt\ a program written in the timed concurrent constraint language
\tccp, which is a logic concurrent constraint language based on the
concurrent constraint paradigm of Saraswat. 
Some inherent notions to \tccp{} processes are 
non-determinism, 
dealing with partial information in states and the monotonic evolution
of the information.

In order to check an \ltl{} property for a process, our approach is based
on the abstract diagnosis technique.
The concluding step of this technique needs to check the
validity of an \ltl{} formula (with constraints) in an effective
way. 

In this paper, we present a decision method for the
validity of temporal logic formulas (with constraints) built by our
abstract diagnosis technique.

\end{abstract}

\section{Introduction}

The \ccp\ paradigm is different from other (concurrent) programming
paradigms mainly due to the notion of store-as-constraint that
replaces the classical store-as-valuation model. It is based on an
underlying constraint system that handles constraints on variables,
thus, it deals with partial information. One challenging
characteristics of the \ccp\ framework is that programs can manifest
non-monotonic behaviors, implying that standard approaches cannot
be directly adapted.  Within this family, \cite{deBoerGM99}
introduced
the \emph{Timed Concurrent Constraint Language} (\tccp)
%     
%\tccp\ 
by adding to the original \ccp\ model the notion of time and the
ability to capture the absence of information.
With these features, one can specify behaviors typical of reactive
systems such as \emph{timeouts} or \emph{preemption} actions.

Modeling and verifying concurrent systems by hand can be very %\emph{really}
complicated.  Thus, the development of automatic formal methods is
essential.
One of the most known technique for formal verification is
model checking, that was originally introduced in
\cite{ClarkeE81,QueilleS82} to automatically check if a finite-state system
satisfies a given property.  It consisted in an exhaustive analysis of the
state-space of the system; thus the state-explosion problem is its main
drawback
and, for this reason, many proposals in the literature try to mitigate it.
\begin{extendedvers}
Some of the more successful ones
are the symbolic approach \cite{BurchCMDH92,HenzingerNSY94,BiereCCSZ03},
%\cite{McMillan93}), 
on-the-fly model checking \cite{Holzmann96}
%,HenzingerKV96
and the abstract interpretation based techniques \cite{ClarkeGL92,Dams96}.
%DillW95,DamsGG95,
The model-checking technique for \tccp\ was first defined in
\cite{FalaschiV06}, and also in this setting (optimized) symbolic and
abstract versions were later defined \cite{AlpuenteFV05,AlpuenteGPV05}.
\end{extendedvers}

All the proposals of model checking have in common that a subset of the
model of the (target) \query\ has to be built, and sometimes the needed
fragment is quite huge.
Our final goal is to define a method that avoids the need to build the
model of a system in order to check the validity of some temporal
property. We propose an extension of the abstract
diagnosis technique of \cite{CominiTV11absdiag} so that the abstract
domain is formed by \ltl{} formulas.
The final step of the method consists in checking whether a given
formula, built from the program (abstract) semantics and the
specification, is valid. 

In this work, we present a decision procedure to check the validity of
such formulas. 
The linear temporal logic used in this work is an adaptation of the
propositional \ltl\ logic to the concurrent constraint framework,
following the ideas of
\cite{PalamidessiV-CP2001,deBoerGM01,deBoerGM02,Valencia05}. It is
expressive enough to represent the abstract semantics of \tccp{} with
much precision and we provide a decision procedure based on the
tableaux method of \cite{GaintzarainHLN08,GaintzarainHLNO09}.
As we show through this paper, the considered logic 
has some differences \wrt\ the classic \ltl{} logic due to the constraint
nature and to the fact that models
for \tccp\ programs have some special characteristics (inherited from
the \ccp\ paradigm).  

\section{The small-step denotational semantics of \tccp}
\label{sec:Sem}

The \tccp{} language \cite{deBoerGM99} is particularly suitable to specify
both reactive and time critical systems.  As the other languages of the
\ccp{} paradigm \cite{Saraswat93}, it is parametric \wrt\ a cylindric
constraint system which
handles the data information of the program in terms of constraints.
The computation progresses as the concurrent and asynchronous activity of
several agents that can (monotonically) accumulate information in a
\emph{store}, or query some information from it.  Briefly, a cylindric
constraint system\footnote{See \cite{deBoerGM99,Saraswat93} for more
details on cylindric constraint systems.} is an algebraic structure
$\CSys=\CCS$ composed of a set of constraints $\CSdom$ such that
$\cpo{\CSdom}{\CSord}$ is a complete algebraic lattice where $\CSmerge$ is
the $\lub$ operator and $\CSfalse$ and $\CStrue$ are respectively the
greatest and the least element of $\CSdom$; $\Var$ is a denumerable set of
variables and $\CShid{}$ 
existentially quantifies variables over constraints.
The \emph{entailment} $\CSimp$ is the inverse of order $\CSord$.

Given a cylindric constraint system $\CSys$ and a set of process symbols
$\Pi$, the syntax of agents is given by the grammar:
\begin{equation*}
    A ::= \askip \mid \atell{c} \mid A \parallel A \mid
    \ahiding{x}{A} \mid
    %\asumask{n}{c}{A}
    \sum_{i=1}^{n}\aask{c_{i}} A
    \mid \anow{c}{A}{A}
    \mid \mgc{p}{x}{}
\end{equation*}
where $c$, $\seq{c}$ are finite constraints in $\CSys$; $p_{/m}\in\Psyms$,
$x\in \Var$ and $\vec{x} \in \Var \times \dots \times \Var$.
A \tccp{} \query\ $P$ is an object of the form $\Qq{D}{A}$, where $A$ is an
agent, called \emph{initial agent}, and $D$ is a set of \emph{process
declarations} of the form $\mgrule{p}{x}{A}$. \begin{extendedvers}(for an agent $A$).\end{extendedvers}

The notion of time is introduced by defining a discrete and global clock.
The $\aask{}$, $\atell{}$ and process call agents take one time-unit to be
executed.

Intuitively, the $\askip$ agent represents the
successful termination of the agent computation.  The $\atell{c}$ agent
adds the constraint $c$ to the current store and stops. It takes one
time-unit, thus the constraint $c$ is visible to other agents from the following
time instant. The store is updated by means of the $\CSmerge$ operator of
the constraint system.  The choice agent $\asumask{n}{c}{A}$ consults the
store and non-deterministically executes (at the following time instant)
one of the agents $A_i$ whose corresponding guard $c_i$ holds in the
current store; otherwise, if no guard is entailed by the store, the agent
suspends.  The agent $\anow{}{}{}\,c\,\mathsf{then}\,A_1\,\mathsf{else}\,A_2$
behaves in the current time instant
like $A_1$ (\resp\ $A_2$) if $c$
is (\resp\ is not) entailed by the store. Entailment is checked
by using the $\CSimp$ operator of the constraint system.
Note that this agent can process
negative information: it can capture when some information is not present
in the store since the agent $A_2$ is executed both when $\neg c$ is
entailed, but also when neither $c$ nor $\neg c$ are entailed.
$A_1 \parallel A_2$ models the parallel composition of $A_1$ and
$A_2$ in terms of maximal parallelism, 
\begin{extendedvers}(in contrast to the interleaving
approach of \ccp{}), \end{extendedvers}
\ie{} all the enabled agents of $A_1$ and $A_2$ are
executed at the same time.
The agent $\ahiding{x}{A}$ is used to
make variable $x$ local to
$A$.  To this end, it uses the $\CShid{}{}$ operator of the constraint system.
Finally, the agent $\mgc{p}{x}{}$ takes from $D$ a declaration of the form
$\mgrule{p}{x}{A}$ and executes $A$ at the following time instant.  
\begin{extendedvers}For the
sake of simplicity,
\end{extendedvers} 
We assume that the set $D$ of declarations is closed
\wrt\ parameter names.

In this work, we
refer to
the denotational concrete semantics defined in
\cite{CominiTV11absdiag} which is fully-abstract \wrt\ the small-step
behavior of \tccp{}.
Due to space limitations, we just introduce intuitively the most relevant
aspects of such semantics and provide a global view of it by means of an
example.
The missing definitions, as well as the proofs of all the results, can be
found in \cite{CominiTV13semTR}.

The denotational semantics of a \tccp{} program consists of a set of
\emph{conditional (timed) traces} that represent, in a compact way, all the
possible behaviors that the program can manifest when
fed with an \emph{input} (initial store). The set of all conditional timed
traces is denoted with $\topC$.
Intuitively, conditional traces can be seen as hypothetical computations
in which, for each time instant, % we have a conditional state where each
a condition represents the information that the global store has to satisfy
in order to proceed to the next time instant.
Briefly, a conditional trace is a (possibly infinite) sequence $t_1\cdots
t_n\cdots$ of \emph{conditional states}, which can be of three forms:
\begin{description}
    \item[conditional store:] 
    a pair $\cs{\eta}{c}$, where $\eta$ is a \emph{condition} and
    $c\in\CSys$ a store;
    
    \item[stuttering:] 
    the construct $\stutt{C}$, with $C \subseteq
    \CSys\setminus\{\CStrue\}$;
    
    \item[end of a process:] 
    the construct $\ed$.
\end{description}
The conditional store $\cs{\eta}{c}$ is used to represent a hypothetical
computation step, where $\eta$ is the condition that the current store must
satisfy in order to make the computation proceed.
$c$ represents the information that is provided  by the program
up to the current time instant.
A \emph{condition} $\eta$ is a pair $\eta=(\eta^{+},\eta^{-})$ where
$\eta^{+} \in \CSys$ and
$\eta^{-} \in \wp(\CSys)$ are 
called positive
and negative condition, \resp.  The positive/negative condition represents
information that a given store must/must not entail,
thus they have to be consistent in the sense that
$\forall c^-\in\eta^-$, $\eta^+\CSnimp c^-$.

\begin{extendedvers}Due to the partial nature of the constraint system, we cannot use
disjunction to represent that \emph{some} given constraints cannot be
entailed, thus we have to use a set of constraints for the negative
condition.
\end{extendedvers}
Conditional states in a conditional trace have to be monotone (\ie\
for each $t_i=\cs{\eta_i}{c_i}$ and $t_j=\cs{\eta_j}{c_j}$ such that
$j\geq i$, $c_j \CSimp c_i$) and consistent (\ie\ each store must not
entail its associated negative conditions).

We abuse in notation and define as $\seqHid{x}{r}$
the sequence resulting by removing from $r$
all the information about the variable $x$.
%
%We distinguish two special classes of conditional traces.
Moreover, $r\in\topC$ is said to be \emph{\closed} if the first condition is
$\cond{\CStrue}{\emptyset}$ and, for each $t_i = \csC{\eta^+_i}{\eta^-_i}{c_i}$ and
$t_{i+1}=\csC{\eta^+_{i+1}}{\eta^-_{i+1}}{c_{i+1}}$, $c_i \CSimp \eta^+_{i+1}$
(each store satisfies the successive condition). 
Moreover, $r$ is \emph{\closed\ \wrt\ $x\in\Vsyms$}
(\emph{$x$-\closed}) if $\seqHid{\Var\setminus \{x\}}{r}$ is \closed.
Intuitively,
%Thus this definition demands that for 
for \closed\ conditional traces, no
additional information (from other agents) is needed in order to
\emph{complete} the computation.  \begin{extendedvers}For $x$-\closed\ conditional traces, the
same happens but only considering information about variable $x$.
 \end{extendedvers}
\begin{example}\label{ex:simple1}
    Let us consider a program with a single process declaration $D \dfn
    \{\prule{p}{y}{A} \}$, where % where the body of the process is defined as 
    \begin{align*}    
        A \dfn \ahiding{x}{}
        (\anow{y=1}{(\aparallel{\atell{x=5}}{\apcall{p}{y}})}{\atell{y=1}})
    \end{align*}

    The semantics of $A$ is graphically represented in \smartref{fig:simple1}.
\begin{extendedvers}    Intuitively, variable $x$ is defined local to
  the body of
    $\apcall{p}{y}$ by means of the operator $\seqHid{}{}$.
\end{extendedvers}
The left branch represents the computation when the information $y=1$
is entailed by the store (the positive condition requires %. The conditional state requires that the constrain 
$y=1$),
thus the information added by the tell
agent ($x=5$) is joined to the store and, in the following state, since it is
invoked a (recursive) process call, we find the interpretation of the process called (represented by the triangle labeled with the interpretation
$\I[]$). 
Process calls do not modify the store when
invoked, but they affect the store from the following time instant.
The right branch is taken
only if $y=1$ does not hold in the current state, then it %is ensured
%that it will 
holds at the following time instant due to the
\atell{} agent.

    \begin{figure}[tp]
        \centering{
        \begin{tikzpicture}[scale=0.45]
        
            \draw (0,0) coordinate (root);
            \draw (root)+(-5,-1) node (one) {$\csC{y=1}{\emptyset}{\CShid{x}{(y=1 \wedge x=5)}}$};
            \draw (root)+(5,-1) node (two) {$\csC{\CStrue}{\{y=1\}}{y=1}$};
            \draw (two)+(0,-1.5) node (three) {$\ed$};
            \draw (one)+(0,-1.3) coordinate (con1);
            %     
            % \draw (two)+(-2.5,-1.5) node (three) {$\csC{y\geq 0}{\emptyset}{y\geq 0}$};
            % \draw (three)+(0,-2) node (four) {$\csC{y\geq 0}{\emptyset}{y\geq 0 \wedge z\leq0}$};
            % \draw (two)+(+3.5,-2.5) node (five) {$\stutt{\{y\geq0\}}$};
            % \draw (four) +(0,-2) node (four_ed) {$\ed$};
            % 
            %arcs
            \draw[-latex] (root) -- (one);
            \draw[-latex] (root) -- (two);
            \draw[-latex] (two) -- (three);
            % \draw[-latex] (near1) -- (near2);
            % \draw[-latex] (out1) -- (out2);
            \draw[-latex] (one) -- (con1);
            % \draw[-latex] (out1) -- (con2);
            % \draw[-latex] (rec1) -- (con3);
            % \draw[-latex] (two) -- (five);
            % \draw[-latex] (three) -- (four);
            % \draw[-latex] (four) -- (four_ed);
        
            % triangle
            \draw (con1.south)+(3.5,-2) coordinate (con13);
            \draw (con1.south)+(-3.5,-2) coordinate (con14);
            \draw (con1.south) -- (con13) -- (con14) -- cycle;
            \draw (con1.south)+(0,-1.4) node (con1call)
            {$\seqHid{x}{\I[](p(y))}$};
        \end{tikzpicture}
        }%
        \caption{Tree representation of $\Aa{A}{\I[]}$  of
        \smartref{ex:simple1}.}
        \label{fig:simple1}
    \end{figure}  

\end{example}

\section{Constraint System Linear Temporal Logic}\label{sec:CLTL}

In this section, we define a variation of the classical Linear
Temporal Logic \cite{MannaP92}.  Following
\cite{PalamidessiV-CP2001,deBoerGM01,deBoerGM02,Valencia05}, the idea
is to replace atomic propositions by constraints of the underlying
constraint system. This logic is the basis for the definition of the
abstract semantics needed in our abstract diagnosis technique.

\begin{definition}[\csltl{} formulas]
    \label{def:tf}
    Given a cylindric constraint system $\CSys$, $c\in\CSys$ and
    $x\in\Var$, formulas of the \csltl{} %\emph{Constraint System Linear Temporal
    Logic over $\CSys$ are defined as: % by using the grammar:
    \begin{equation*}
        \phi ::= \Ltrue \mid \Lfalse \mid c \mid \Lneg{\phi} \mid
        \Lconj{\phi}{\phi} \mid \Lhid{x}{\phi} \mid \Lnext{\phi} \mid
        \Luntil{\phi}{\phi}.
    \end{equation*}

    We denote with $\csltl$ the set of all temporal formulas over $\CSys$.
\end{definition}
The formulas $\Ltrue$, $\Lfalse$, $ \Lneg{\phi}$, and
$\Lconj{\phi_1}{\phi_2}$ have the classical logical meaning.  The atomic
formula $c\in\CSys$ states that $c$ has to be entailed by the current
store.
$\Lhid{x}{\phi}$ is the existential quantification over the set of
variables $\Var$.  $\Lnext{\phi}$ states that $\phi$ holds at the next time
instant, while $\Luntil{\phi_1}{\phi_2}$ states that $\phi_2$ eventually
holds and in all previous instants $\phi_1$ holds.
In the sequel, % (as usual) 
we use $\Ldisj{\phi_1}{\phi_2}$ as a shorthand for
$\Lneg{(\Lconj{\Lneg{\phi_1}}{\Lneg{\phi_2}})}$; $\Limpl{\phi_1}{\phi_2}$ for
$\Ldisj{\Lneg{\phi_1}}{\phi_2}$; %$\Liff{\phi_1}{\phi_2}$ for
% $\Lconj{\Limpl{\phi_1}{\phi_2}}{\Limpl{\phi_2}{\phi_1}}$;
$\Leventually{\phi}$ for $\Luntil{\Ltrue}{\phi}$ and $\Lalways{\phi}$ for
$\Lneg{\Leventually{\Lneg{\phi}}}$.
\begin{extendedvers}
    $\Lwuntil{\phi_1}{\phi_2}$ for
    $\Ldisj{(\Luntil{\phi_1}{\phi_2})}{\Lalways{\phi_1}}$.
    $\Leventually{\phi}$ holds if at some point in the future $\phi$ is true,
    and $\Lalways{\phi}$ holds if $\phi$ holds in the current instant and
    always in the future.
\end{extendedvers}
A \emph{constraint formula} is an atomic formulas $c$ or its negation
$\Lneg{c}$.  Formulas of the form $\Lnext{\phi}$ and
$\Lneg{\Lnext{\phi}}$ are called \emph{next} formulas.
Constraint and next
formulas are said to be \emph{elementary} formulas.
Finally,
formulas of the form $\Luntil{\phi_1}{\phi_2}$ %, $\Lalways{\phi}$
(or $\Leventually{\phi}$ or $\Lneg{(\Lalways{\phi})}$) are called \emph{eventualities}.
    
The truth of a formula $\phi \in \csltl$ is defined \wrt{}
a trace $r \in\topC$.
As usually done in the context of temporal logics, we define the
satisfaction relation $\asat{}{}$ only for infinite conditional traces.
We implicitly transform finite traces (which end in $\ed$) by
replicating the last store infinite times.  
\begin{extendedvers}Namely, the trace
$\csC{\eta_1^+}{\eta_1^-}{c_1}\dots \csC{\eta_n^+}{\eta_n^-}{c_n} \cdot
\ed$ becomes $\csC{\eta_1^+}{\eta_1^-}{c_1}\dots
\csC{\eta_n^+}{\eta_n^-}{c_n} \cdot \replicate{
\csC{c_n}{\emptyset}{c_n} }$, while $\ed$ becomes
$\replicate{\csC{\CStrue}{\emptyset}{\CStrue}}$.
\end{extendedvers}

\begin{definition}%[Satisfaction relation for conditional traces]
    \label{def:FtoM}
    \label{def:asat}
    
    The \emph{semantics} of $\phi\in\Fdom$ is given by $\FtoM{} \colon
    \Fdom \rightarrow \C$ defined as
    \begin{equation}\label{eq:FtoM}
        \FtoM{\phi} \dfn \lubC{ \set*{ r\in \topC }{ \asat{r}{\phi} } }{},
    \end{equation}
    where, for each $\phi, \phi_1, \phi_2 \in \csltl$, $c \in \CSys$ and
    $r\in\topC$, the satisfaction relation $\asat{}{}$ is defined as:
    \begin{subequations}
        \label{eq:asat_rel}
        \begin{align}
            &\asat{r}{\Ltrue}\\
            &\notasat{r}{\Lfalse}\\
            &\asat{\csC{\eta^+}{\eta^-}{d} \cdot r'}{c}&&\text{iff }
            \eta^+ \CSimp c
            \label{eq:asatKnowCS}\\
            &\asat{\stutt{\eta^-} \cdot r'}{c} &&\text{iff } \forall
            d^{-}\in \eta^- .\, c \CSnimp d^{-} \text{ and } \asat{r'}{c}
            \label{eq:asatKnowST}\\
            &\asat{r}{\Lneg{\phi}} &&\text{iff } \notasat{r}{\phi}
            \label{eq:asatNeg} \\
            &\asat{r}{\Lconj{\phi_1}{\phi_2}} &&\text{iff } \asat{r}{\phi_1}
            \text{ and } \asat{r}{\phi_2} \label{eq:asatConj} \\
            &\asat{r}{\Lhid{x}{\phi}} &&\text{iff }
            \begin{aligned}[t]
                &\text{%it exists 
$\exists r'\; $s.t.$\seqHid{x}{r'}=\seqHid{x}{r}$, } %\\
                %&
\text{$r'$ $x$-\closed,
                $\asat{r'}{\phi}$} 
            \end{aligned}\label{eq:asatHid} \\
            & \asat{t \cdot r}{\Lnext{\phi}} &&\text{iff}\ \asat{r}{\phi}
            \label{eq:asatNext} \\
            & \asat{r}{\Luntil{\phi_1}{\phi_2}} &&\text{iff }            
            \exists i\geq
            1 .\, \forall j < i .\ \asat{r^i}{\phi_2}\ \text{and}\ \asat{r^j}{\phi_1}
            \label{eq:asatUntil}
        \end{align}
    \end{subequations}
        We abuse of notation and we extend %the notion of 
$\asat{}{}$ to sets
        of formulas: % in the following way
%        \begin{equation}\label{eq:asat_set}
            $\asat{r}{\Phi} \iff\ \forall \phi\in\Phi.\ \asat{r}{\phi}$
%        \end{equation}
%        
        A formula $\phi$ is said to be \emph{satisfiable} if there exists
        $r\in\topC$ such that $\asat{r}{\phi}$, while it is said to be
        \emph{valid} if, for all $r\in\topC$, $\asat{r}{\phi}$.
 \end{definition}

\begin{extendedvers}
Let us show some temporal properties that can be expressed by \csltl{}
formulas.
\end{extendedvers}
\begin{example}
    The formula $\Lalways(\Limpl{x>0}{\Lnext{z>1}})$
    expresses that forever, %  always in the future,
    whenever $x>0$ is entailed by the %current constraint 
    store, then
    $z>1$ is entailed at the following time instant.
\begin{extendedvers}    
    The formula $\Lhid{x}{(\Leventually(\Lconj{y=1}{x=5}))}$
    expresses that, eventually in the future, $y=1$ is
    entailed by the global constraint store
    and, at the same time instant, there exists a local variable $x$ such that $x=5$
    is locally entailed. % by the local constraint store.
\end{extendedvers}
\end{example}    

\section{Abstract diagnosis of temporal properties}
\label{sec:abs-diag}

Abstract diagnosis is a semantic based method to identify bugs in
programs.  It was originally defined for logic programming
\cite{CominiLMV96a} and then extended to other paradigms
\cite{AlpuenteCEFL02,BacciC10absdiag,FalaschiOPV07,CominiTV11absdiag}.
This technique is based on the definition of an abstract semantics for
the program which must be a sound approximation of its behavior.
Then, given an
abstract specification $\SF$ of the expected behavior of the program, the
abstract diagnosis technique 
automatically detects the errors in the program by checking if the
result of one
computation of the semantics evaluation function
(where the procedure calls are interpreted over $\SF$)
is \lq\lq contained\rq\rq{} in the specification itself.

A first approach to the abstract diagnosis of \tccp{}
was presented in \cite{CominiTV11absdiag} by using as specifications
sets of abstract traces. The main drawback of that proposal was that specifications
were given in terms of traces, thus they can be tedious to write. In order
to overcome that problem, we have defined an abstract semantics \begin{extendedvers}(a
semantics evaluation function)\end{extendedvers} in terms of \csltl{} formulas. This
allows us to express the intended behavior in a more compact way, by means of a \csltl{} formula.

The semantics evaluation function $\FAa{A}{}$, given an agent $A$ and an
interpretation $\FI$ (for the process symbols of $A$), builds a \csltl{}
formula representing a correct approximation of the small-step
behavior of $A$.\begin{extendedvers}\footnote{Since it may help the reader to follow the
upcoming discussions, we have included in the appendix the definition of
$\FAa{}{}$ although we know that some operators and notation remain
undefined. Since the scope of this paper is not to convince the reader about
the correctness %and completeness 
of that semantics, we beg him
to accept it as it is.}
\end{extendedvers}
The semantics of the declarations is given in terms of the fixpoint of a
semantics evaluation function $\FDd{D}{}$
which associates to each procedure declaration $\mgc{p}{x}{}$ the logic disjunction
of the abstract semantics of every agent $A$ such that
$\mgrule{p}{x}{A}$ belongs to the declaration $D$
(\ie{} $\FDd{\P}{\FI} (\mgc{p}{x}{}) \dfn \lubF{_{\mgrule{p}{x}{A} \in D}
\FAa{A}{\FI}} {}$).

The \emph{abstract diagnosis} technique determines exactly the
``originating'' symptoms and, in the case of incorrectness, the faulty
\progrule\ in the program.  This is captured by the definitions of
\emph{abstractly incorrect \progrule} and \emph{abstract uncovered
  element}.  Informally, a process declaration $\P$ is abstractly
incorrect if it derives a wrong abstract element $\phi_{t}\in\csltl$
from the intended semantics $\SF$.  Dually, $\phi_{t}$ is uncovered if
the declarations cannot derive it from the intended semantics.

We show
here the main result of abstract diagnosis, which determines the form
of formulas that we need to check for validity.
\begin{theorem}%[Correctness and weak completeness]
    \label{th:ab.corr-compl}
    
    Consider a set of declarations $\P$ and an abstract specification $\SF$.
    \begin{enumerate}
        
        \item\label{pt:ab.correct} If there are no abstractly incorrect
        process declarations in $\P$ (\ie\ $\FDd{\P}{\SF} \leqF \SF$),
        then $\P$ is partially correct \wrt\ $\SF$ (the small-step denotational
        semantics of $\P$ is \lq\lq contained \rq\rq in the semantics of $\SF$).
       
        \item\label{pt:ab.complete2} Let $\P$ be partially correct \wrt\
        $\SF$.  If $\P$ has abstract uncovered elements then $\P$ is not
        complete.
    \end{enumerate}
\end{theorem}
Therefore, in order to check partial correctness of a program, it is
sufficient to check the implication $\FDd{\P}{\SF} \leqF \SF$. 

Because of the approximation, it can happen that a (concretely) correct
declaration is abstractly incorrect.  Hence, abstract incorrect declarations
are in general just a warning about a possible source of errors.
However, an abstract correct declaration cannot contain an error;
therefore, no (manual) inspection is needed for declarations which are not
signalled.  Moreover, it happens that %as shown by the following
                                %theorem, 
all concrete
errors---that are ``visible''---are detected, as they lead to an
abstract incorrectness or abstract uncovered. 

\begin{extendedvers}
Let us now illustrate how the technique works by means of a simple
example. % an
\end{extendedvers}
\begin{example}\label{ex:simpleDiag} 
    Assume we need to check that the program in \smartref{ex:simple1}
    satisfies that the constraint $y=1$ is eventually
    entailed by the store.
%    that the information $y=1$ is eventually entailed by the store.
    The intended specification for the process $p$ corresponding to
    this property is $\SF (\apcall{p}{y}) \dfn \Leventually{(y=1)}$.
    The \csltl-semantics $\FDd{}{}$ 
for $\apcall{p}{y}$ with the
    given specification as interpretation is
    \begin{align*}
        \FDd{D}{\SF}(\apcall{p}{y}) = \Lhid{x}{\big( \Ldisj{(\Lconj{y=1}{\Lconj{\Lnext{x=5}}{\Lnext{(\Leventually{y=1})}}})}
        {(\Lconj{\Lneg{y=1}}{\Lnext{y=1}})} \big)}
    \end{align*}
    
    Note that the resulting formula has a clear correspondence with
    the behavior of the program. We have two disjuncts, one for each
    branch of the conditional in the body of the
    declaration. Moreover, since the conditional agent is in the scope
    of a local variable $x$, both disjuncts are enclosed within an
    existential quantification. The computation of the
    \csltl-semantics is compositional, based on the structure of the
    program. The first disjunct corresponds to the case when the guard
    of the conditional agent is satisfied ($y=1$), thus in the
    following time instant two things
    happen: \begin{extendedvers}$x=5$ is added to the store,
      thus\end{extendedvers} $x=5$ is entailed, and also is entailed
    the interpretation for the process call $p$ because a recursive
    call is run. This illustrates how the intended specification is
    used as the interpretation of process calls.

    Following \smartref{th:ab.corr-compl}, to check whether the
    process $\apcall{p}{y}$ satisfies the property, %specification
    it is sufficient to show that \begin{extendedvers}the \csltl{} formula\end{extendedvers}
    $\FDd{D}{\SF}(\apcall{p}{y}) \leqF \SF (\apcall{p}{y})$
    is valid.
\end{example}

The fact that our technique is based on abstract diagnosis becomes
explicit when two situations occur simultaneously. When the process
that is being analyzed has more than one fixpoint (this essentially
happens when $\P$ contains a loop which does not produce contributes
at all) and the specification is an eventuallity that is not
inconsistent with the program behavior, %, it becomes explicit that our
then it may happen that the actual behaviour does not model a
specification $\SF$, which is a non-least fixpoint of $\FDd{\P}{}$,
but we do not detect abstractly incorrect declarations since $\SF$ is
a fixpoint. This is natural in the context of the abstract diagnosis
technique: what we are proving is that,
% 
%However, 
if $\SF (\mgc{p}{x}{})$ is assumed to hold for each process
$\mgc{p}{x}{}$ defined in $\D$ and $\Limpl{\FDd{ \D }{\SF}} {{\SF}}$,
then \emph{the program} $\F[]{\P}$ satisfies $\SF$.

In any other situation, a positive result of the abstract diagnosis
corresponds to proving that the program \emph{strongly} satisfies the
temporal property. Also if we provide a specification that is in
contradiction with the actual behavior of the process, our technique
answers as expected in a verification context.

\section{An automatic decision procedure for \csltl}\label{sec:decision}

In order to make our abstract diagnosis approach effective,
we need to define an automatic decision procedure to
check the validity of the \csltl{}
formulas that show up when checking a property.
In particular, we need to handle \csltl{} formulas of the form  
$\Limpl{\psi}{\phi}$, where
$\psi$ corresponds to the computed approximated behavior of the program,
and $\phi$ is the abstract intended behavior
of the process. 

\Laura[non sfruttiamo la particolare forma di
$\phi$ e $\psi$
quindi questa grammatica si potrebbe togliere]{

Formally, $\phi$ and $\psi$
are defined by the following grammars.
\begin{align*}
    \phi &\dfn \Ltrue \mid \Lfalse \mid c \mid \Lneg{\phi} \mid
        \Lconj{\phi}{\phi} \mid \Lnext{\phi} \mid
        \Luntil{\phi}{\phi}\\
    \psi &\dfn c \mid \Lneg{c} \mid \Lnext{\psi} \mid \Lconj{\psi_1}{\psi_2} \mid
    \Ldisj{\psi_1}{\psi_2} \mid \Lhid{x}{\psi} \mid \Lnext{\phi}
\end{align*}

Due to the definition of the abstract semantics \begin{extendedvers}(see in particular the
definition of
$\FAa{}{}$ in appendix)\end{extendedvers}, $\psi$ cannot be an arbitrary \csltl\
formula. We know
that the until operator can occur only in the scope of a next
operator (it can show up only thanks to the intended specification in
a process call, whose execution has a delay of one time instant).
It
also happens that negation can be applied to arbitrary formulas if
they are within a next operator, otherwise it can be applied only to constraints. 
}

We impose a restriction on the specification $\phi$: we do not allow the use
of existential quantifications. Actually, this restriction is quite
natural in our context since, in general, we are interested in proving properties related
to the \emph{visible} behavior of the program, not to the local variables.
In contrast, negation can be applied to any formula $\phi$
(not only to constraints).

In this section, we extend the tableau construction for Propositional
LTL (\pltl) of
\cite{GaintzarainHLN08,GaintzarainHLNO09}  in order to deal with
\csltl\ formulas.
We need to adapt the method to our context due to three issues:
\begin{enumerate}
    \item The structures on which the logic is interpreted are
    different. In our case, traces (sequences of states) are monotonic,
    meaning that the information in each state always increases. 
    \item The logic itself is a bit different from \pltl\ since propositions are replaced by constraints in $\CSys$.
    \item We have to handle existential quantification over variables of the
    underlying constraint system. This does not mean that we are dealing
    with a first-order logic as will become clear later
\end{enumerate}

In the following, we first present the basic rules that are used during the
construction of the tree associated to the tableau. Then we present
the algorithm that implements the process of construction of the tree.

\subsection{Basic rules for a \csltl{} tableau}\label{subsec:tab_rules}

Classic tableaux algorithms are based on the systematic construction of
a graph which is used to check the satisfiability of the formula. In
\cite{GaintzarainHLN08,GaintzarainHLNO09}, the authors present an algorithm that does
not need to use auxiliary structures such as graphs to decide about the
satisfaction of the formula, and this makes this approach more suitable
for automatization.
 
A tableau procedure is defined by means of rules that build a tree
whose nodes are labeled with sets of formulas. If all
branches of the tree are \emph{closed}, then the formula has no
models. Otherwise, we can obtain a model that satisfies the formula from
the open branches.
Let us introduce the basic rules for the \csltl{} case. As usual, we present
just the minimal set of rules.

A tableau rule is applied to a node $n$ labeled with the set of formulas $\Tlabel{n}$.
Each rule application requires a previous selection of a formula
$\phi$ from $\Tlabel{n}$.
We call context to the set of formulas $\Tlabel{n}\setminus \{\phi\}$ and we denote it with $\Gamma$. 
Conjunctions are $\alpha$-formulas and disjunctions
$\beta$-formulas. \smartref{fig:alpha_rules} presents the rules for
$\alpha-$ and $\beta-$formulas.

\begin{figure*}
    \begin{center}
    \mbox{ 
\begin{tabular}{ | c | c | c | }
    \hline
   & $\alpha$ & $\Arule{\alpha}$\\ \hline
   \ R1\ \setlabel{R1}\label{rule:neg} & $\Lneg{\Lneg{\phi}}$ & $\set{\phi}{}$ \\ \hline
   R2\setlabel{R2}\label{rule:conj} & $\Lconj{\phi_1}{\phi_2}$ & $\set{\phi_1,\phi_2}{}$ \\ \hline
   %R3\setlabel{R3}\label{rule:always} & $\Lalways{\phi}$ & $\phi,\Lnext{\Lalways{\phi}}$ \\ \hline
\end{tabular}
%\caption{$\alpha$-formulas rules}
\hspace{3mm} 
\begin{tabular}{ | c | c | c | c | }
    \hline
   & $\beta$ & $\Brulel{\beta}$ & $\Bruler{\beta}$\\ \hline
    \ R3 \setlabel{R3}\label{rule:negconj} & $\Lneg{(\Lconj{\phi_1}{\phi_2})}$ & $\set{\Lneg{\phi_1}}{}$ & $\set{\Lneg{\phi_2}}{}$ \\ \hline
    \ R4 \setlabel{R4}\label{rule:neguntil} & $\Lneg{(\Luntil{\phi_1}{\phi_2})}$ & $\set{\Lneg{\phi_1},\Lneg{\phi_2}}{}$
    & $\set{\phi_1,\Lneg{\phi_2},\Lneg{\Lnext{(\Luntil{\phi_1}{\phi_2})}}}{}$ \\ \hline
   \ R5 \setlabel{R5}\label{rule:until} & $\Luntil{\phi_1}{\phi_2}$ & $\set{\phi_2}{}$ & $\set{\phi_1,\Lneg{\phi_2},\Lnext{(\Luntil{\phi_1}{\phi_2})}}{}$ \\ \hline
   \ R6 \setlabel{R6}\label{rule:dist_until} & $\Luntil{\phi_1}{\phi_2}$ &
   $\set{\phi_2}{}$ & $\set{\phi_1,\Lneg{\phi_2},\Lnext{(\Luntil{(\Lconj{\cntx{\Gamma}}{\phi_1})}{\phi_2})}}{}$ \\ \hline
\end{tabular}
}
\end{center}
\caption{$\alpha$- and $\beta$-formulas rules}\label{fig:alpha_rules}\label{fig:beta_rules}
\end{figure*}

Tables in \smartref{fig:alpha_rules} are interpreted as follows. Each row in a table
represents a rule. Each time that an $\alpha-$rule is
applied to a node of the tree, a formula of the node matching the
pattern in column $\alpha$ is replaced in a child node by the
corresponding $\Arule{\alpha}$. For the $\beta$-rules, two children nodes
are generated, one for each column $\Brulel{\beta}$ and $\Bruler{\beta}$.

Almost all the rules are standard. However, \smartref{rule:dist_until} uses the
so-called context $\cntx{\Gamma}$, which is defined as $\cntx{\Gamma} \dfn
\lubF{}{}_{\gamma \in \Gamma} \Lneg{\gamma}$.  The use of contexts is the mechanism to detect the
loops where no formula changes, thus allowing to mark branches containing eventually formulas as \emph{open}. This
kind of rules %The Special rules \smartref{rule:dist_until} and \smartref{rule:dist_eventually}
were first used in \cite{GaintzarainHLNO07}. \begin{extendedvers}The idea is that, by
using contexts, loops where no formula changes are \emph{discarded}
since they cannot close a branch.
\end{extendedvers}

Note that there is no rule defined for the $\Lnext{}$ operator. In
fact, the $\nextop{\Phi}$ function transforms a set of
elementary formulas $\Phi$ into another:
$\nextop{\Phi} \dfn \set{\phi}{\Lnext{\phi} \in \Phi} \cup
\set{\Lneg{\phi}}{\Lneg{\Lnext{\phi}} \in \Phi} \cup \set{c}{c\in\Phi,
c\in \CSys }$. This operator is different from the corresponding one
of \pltl{} in that, in addition to keeping the internal formula of the
next formulas, it also \emph{passes} the constraints that are entailed
at the current time instant  to the following one. This makes sense
for \tccp{} computations since, as already mentioned, the store in a
computation is monotonic, thus no information can be removed and it
happens that, always, $c$ implies $\Lnext{c}$.

The next operator is a key notion in the kind of tableaux defined in
\cite{GaintzarainHLN08,GaintzarainHLNO09}. This operator allows one to
identify \emph{stages} in a tableau which represent time instants in
the model. \begin{extendedvers}Some additional checks (explained in the following
sections) allow to avoid the use of auxiliary graph representations for
determining satisfiability/unsatisfiability of formulas.
\end{extendedvers}

We show that $\alpha$- and $\beta$-formulas rules and the $\nextop{}$ operator
preserve the satisfiability of a set of formulas. 

\begin{lemma}\label{lem:step_corr}
    Given a set of formulas $\Phi$, an $\alpha$-formula $\alpha$
    and a $\beta$-formula $\beta$:
    \begin{enumerate}
        \item $\Phi \cup \{ \alpha \}$ is satisfiable $\Leftrightarrow$ $\Phi \cup  \Arule{\alpha} $ is satisfiable;
        \item $\Phi \cup \{ \beta \}$ is satisfiable $\Leftrightarrow$ $\Phi \cup  \Brulel{\beta} $
        or $\Phi \cup \Bruler{\beta} $ is satisfiable;
        \item if $\Phi$ is a set of elementary formulas,
        $\Phi$ is satisfiable $\Leftrightarrow$ $\nextop{\Phi}$ is satisfiable;
    \end{enumerate}    
\end{lemma}

A second main difference \wrt{} the \pltl\ case regards the
existential quantification.
The \csltl{} existential quantification \begin{extendedvers}does not correspond
to the first-order logic one. It \end{extendedvers}is introduced to model information
about local variables, thus, the formula $\Lhid{x}{\phi}$
can be seen as the formula $\phi$ where the information about $x$
is local.

We define a specific rule for the $\Lhid{}$ case: %. In particular, 
%During the construction of a tableau, 
when the selected formula of a
given node is of the form $\Lhid{x}{\phi}$, %we 
it is created a node, %This rule generates a 
child of $n$, whose labeling is that of $n$ except that %where 
the formula $\Lhid{x}{\phi}$
is replaced by $\phi$. 
Correctness of this rule derives from the following lemma, which shows that $\Lhid{x}{\phi}$ and $\phi$ are equi-satisfiable. 
\begin{lemma}\label{lem:exists_corr}
    Let $\phi\in\csltl$, $\Lhid{x}{\phi}$ is satisfiable $\iff$
    $\phi$ satisfiable.
\end{lemma}

\begin{extendedvers}
\begin{corollary}\label{cor:exists_corr}
    Let $\Phi\subseteq \csltl$ such that $x\in\Var$
    does not appear in $\Phi$
    and let $\phi\in\csltl$.
    Then, $\Phi \cup \{\Lhid{x}{\phi}\}$ is satisfiable $\iff$
    $\Phi \cup \{\phi\}$ is satisfiable.
\end{corollary}    

\begin{extendedvers}
\begin{proof}
    Follows directly from \smartref{lem:exists_corr}.
    $x$ does not appear in $\Phi$, thus the local variable $x$ of
    $\phi$ is independent from any other variable in $\Phi$.
\end{proof}    
\end{extendedvers}

It can be noticed that the fact that $x$ cannot appear in 
$\Phi$ is not a real restriction since it is possible to perform a renaming
in order to apply safely the $\Lhid{}$ elimination
without incurring in variable names crushes.
\end{extendedvers}
\subsection{Semantic \csltl{} tableaux}

In this section, we present the notion of tableau %method 
for
our \csltl{} formulas
following the ideas
of \cite{GaintzarainHLN08,GaintzarainHLNO09}. Since we borrow some
definitions and notions from that %previous 
work, in this section we
skip some formal definitions. 

A tableau $\Tname$ for a set of formulas $\Phi$
is a tree-like structure where each node $n$ is labeled with a set of
\csltl{} formulas $\Tlabel{n}$.
The root is labeled with 
the set of formulas $\Phi$
% singleton set $\{ \phi \}$ for the formula $\phi$
whose satisfiability/unsatisfiability is needed to check; Then, children of
nodes are the result of 
applying the basic rules of \smartref{subsec:tab_rules}. The algorithm
in which these nodes are built is given in the following subsection.
Nodes with no children are called \emph{leaf} nodes.

\begin{definition}[\csltl{} tableau]\label{def:tableau}
    A \csltl{} tableau for a finite set of formulas $\Phi$ is a tuple
    $\Tname = \Ttabl$ such that:
    \begin{enumerate}
        \item $\Tnodes$ is a finite non-empty set of nodes;
        \item $\Tinit \in \Tnodes$ is the initial node;
        \item $\Tlabel{}: \Tnodes \rightarrow \wp(\csltl)$
        is the labeling function that associates to each node the formulas which are
        true in that node; the initial node is labeled with $\Phi$;
\begin{extendedvers}        (\ie\ $\Tlabel{\Tinit} = \Phi$);\end{extendedvers}
        \item $\Tbranches$ is the set of branches such that exactly one of the following points
        holds for every $b = n_0, \dots,n_i,n_{i+1},\dots,n_k \in \Tbranches$ and every $0 \leq i < k$:
            \begin{enumerate}
                \item $\Tlabel{n_{i+1}} = \{\Arule{\alpha}\} \cup \Tlabel{n_{i}}\setminus \{\alpha\}$
                for an $\alpha$-formula $\alpha\in \Tlabel{n_{i}}$;
                \item $\Tlabel{n_{i+1}} = \{\Brulel{\beta}\} \cup \Tlabel{n_{i}} \setminus \{\beta\}$
                and there exists \begin{extendedvers}another branch on the form\end{extendedvers} $b' = n_0, \dots,n_i,n'_{i+1},\dots,n'_k$
                such that $\Tlabel{n'_{i+1}} = \{\Bruler{\beta}\} \cup \Tlabel{n_{i}} \setminus \{\beta\}$
                for a $\beta$-formula $\beta \in \Tlabel{n_{i}}$;
                \item $\Tlabel{n_{i+1}} = \{ \phi'\} \cup \Tlabel{n_{i}} \setminus \{\Lhid{x}{\phi'}\}$
                for %for an existential quantified formula 
$\Lhid{x}{\phi'} \in \Tlabel{n_{i}}$;
                \item $\Tlabel{n_{i+1}} = \nextop{\Tlabel{n_{i}}}$ if
                $\Tlabel{n_{i}}$ contains only
                elementary formulas.
                %by constraint and next formulas.
            \end{enumerate}    
    \end{enumerate}
\end{definition}

A branch $b\in\Tbranches$ is \begin{extendedvers}said to be\end{extendedvers} maximal if it is not a proper prefix of another branch in $B$.

\begin{definition}
    A node in the tableau is \emph{inconsistent} if it contains 
    \begin{itemize} 
        \item a couple of formulas
        $\phi, \Lneg{\phi}$, or 
        \item the formula $\Lfalse$, or
        \item a couple of constraint formulas $c, \Lneg{c'}$ such that %$c,c'\in\CSys$
        $c\CSimp c'$.
    \end{itemize}
\end{definition}

The last condition for inconsistence of a node is particular to the
\ccp{} context. Since we are dealing with constraints that model
partial information, it is possible to have an \emph{implicit}
inconsistence, in the sense that we need the entailment relation to
detect it.

An inconsistent node does not accept any rule application.
When a branch contains an inconsistent node, it is said to be closed, otherwise it is % said to be
open.

By \smartref{lem:step_corr} and by \smartref{def:tableau}, it can be noticed that every closed branch contains
only unsatisfiable sets of formulas. Open branches are not necessarily satisfiable since they could be
prefixes of a closed one.

Similarly to the \pltl\ case, 
%As we will formalize later, in a tableau $\Tname[\Phi]$
it exists only a finite number of different labels in a
tableau. Thus, %if there exists 
an infinite 
branch $b = n_0,n_1,\dots n_k\ldots$ %, it 
contains a cycle\begin{extendedvers}(\ie\
contains infinitely many repetition of nodes with the same label)\end{extendedvers}.
These branches are called \emph{cyclic branches} and can be finitely
represented as $\Tpath{b} = n_0, n_1, \dots, n_j, (n_{j+1},\dots,
n_k)^{\omega}$ when $\Tlabel{n_k} = \Tlabel{n_j}$ for $0 \leq j <k$.

Every branch of a tableau is divided into stages.
A \emph{stage} is a sequence of consecutive nodes between two consecutive
applications of the $\nextop{}$ operator.
We abuse of notation and we say that the labeling of a stage $s$ is
the
labeling of each node in that stage %by extending the labeling function from nodes to stages, given a stage $s$
($\Tlabel{s} = \bigcup_{n \in s} \Tlabel{n}$). 
\begin{extendedvers}It can be noticed that if $b$ contains a cyclic sequence of nodes, then $\Tstages{b}$
is a cyclic sequence of stages.\end{extendedvers}
Moreover, %intuitively, 
a stage $s$ is \emph{saturated} if no $\alpha$-, $\beta$- or
hiding rule can be applied to any of its nodes. 

We borrow from \cite{GaintzarainHLN08,GaintzarainHLNO09} the
characterization of \emph{fulfilled} eventually formula in a path
of the tableau, namely when it is satisfied. We
say that, when an eventually formula $\Luntil{\phi_1}{\phi_2}$ belongs to the labeling of a
stage $s$ in a path, it is fulfilled if there exists a subsequent
stage $s'$ such that $\phi_2\in\Tlabel{n'}$. 
A sequence of stages $S$ is fulfilling if all the eventually formulas
it its labeling are
fulfilled in $S$ and a branch $b$ is fulfilling if the sequences of stages in its
paths are fulfilling.

Finally, an open branch is expanded if it is fulfilling and all its
stages are saturated. These notions are needed to formalize the
tableau construction since only branches that are non-expanded and
open are selected to be further developed.

\begin{definition}[expanded \csltl{} tableau \cite{GaintzarainHLN08,GaintzarainHLNO09}]
    A tableau is called expanded if every branch is expanded or closed.
    An expanded tableau is closed if every branch ends in an inconsistent node,
    otherwise it is open.
\end{definition}

\subsection{A systematic \csltl{} tableaux construction}

\smartref{def:alg_tabl} presents
 the algorithm to automatically build
an expanded \csltl{} tableau (called systematic tableau %following
\cite{GaintzarainHLN08,GaintzarainHLNO09}) for a given set of formulas
$\Phi$. 

The construction consists in selecting at each step a non-expanded branch 
to be enlarge by using $\alpha$ or $\beta$ rules
or $\Lhid{}$ elimination.
When none of these can be applied,
the $\nextop{}$
operator is used to pass to the next stage.

When dealing with eventualities, to determine which rule
\ref{rule:until} or \ref{rule:dist_until} has to be applied in a node,
it is necessary to \emph{distinguish} the eventuality %. The idea is to
%mark which is the eventuality that 
that is being unfolded in the path. In
this way, the rule \ref{rule:dist_until} is applied only to
\emph{distinguished} eventualities when selected; when
the selected eventuality is not the distinguished one, then rule
\ref{rule:until} is used. If a node does not contain any distinguished
eventuality, then the algorithm distinguishes one of them and rule
\ref{rule:dist_until} is chosen to be applied to it. Each node of the
tableau has at most one distinguished eventuality.

The algorithm marks \begin{extendedvers}nodes when they cannot be further processed. In
particular, \end{extendedvers}a node \begin{extendedvers}is marked\end{extendedvers} as \emph{closed} when it is inconsistent
and \begin{extendedvers}is marked\end{extendedvers}
as \emph{open} when
it contains just constraint formulas or
when it is the last node of an expanded branch (all the eventualities
in the path are fulfilled).

\begin{definition}\label{def:alg_tabl} 

Given as input a finite set of formulas $\Phi$, \begin{extendedvers}the following
algorithm construct the systematic tableau $\Tname$.
\end{extendedvers}
the algorithm repeatedly selects an unmarked leaf node $l$ labelled
with the set of formulas $\Tlabel{l}$ and applies, in order, one of the
points shown below. 

\begin{enumerate}
    \item\label{pt:alg_closed} If $l$ is an inconsistent node, then mark it as closed ($\closedleaf$).
    \item\label{pt:alg_open} If $\Tlabel{l}$ is a set of constraint formulas, mark $l$ as open ($\openleaf$).
    \item\label{pt:alg_cycle} If $\Tlabel{l} = \Tlabel{l'}$ for $l'$ ancestor of $l$,
    take the oldest ancestor $l''$ of $l$ that is labeled with $\Tlabel{l}$
    and check if each eventuality in the path between $l''$ and $l$ is fulfilled
    in such path. If they are all fulfilled, then mark $l$ as open ($\openleaf$).
    \item 
      Otherwise, choose $\phi \in
    \Tlabel{l}$ such that $\phi$ is not a \emph{next} formula. Then,
    \begin{itemize}
        \item\label{pt:alg_alpha} if $\phi$ is an $\alpha$-formula (let $\phi = \alpha$), create a new node $l'$ as a child of $l$
        and label it as $\Tlabel{l'} = (\Tlabel{l}\setminus
        \{\alpha\})\cup{\Arule{\alpha}}$ by using the $\alpha-$rules
        % corresponding rule 
in
        \smartref{fig:alpha_rules},
        \item\label{pt:alg_beta} if $\phi$ is a $\beta$-formula (let $\phi = \beta$), create two new nodes $l'$ and $l''$ as children of $l$
        and label them as $\Tlabel{l'} = (\Tlabel{l}\setminus \{\beta\})\cup{\Brulel{\beta}}$
        and $\Tlabel{l''} = (\Tlabel{l}\setminus \{\beta\})\cup{\Bruler{\beta}}$ by using the corresponding rule in
        \smartref{fig:beta_rules}. Moreover, 
        if $\beta$ is an eventuality, we have three possible cases:
        \begin{itemize}
            \item if $\beta$ is the distinguished eventuality in $\Tlabel{l}$, then apply
            \smartref{rule:dist_until}
            %(or \smartref{rule:dist_eventually})
            to $\beta$
            and distinguish the formula inside the \emph{next} formula in $\Bruler{\beta}$;
            \item if $\beta$ is not distinguished,
            but there is another distinguished formula, then apply
            \smartref{rule:until}
            %(or \smartref{rule:eventually})
            to $\beta$ and maintain the existing distinguished formula in
            $\Brulel{\beta}$ and $\Bruler{\beta}$;
            \item otherwise, distinguish $\beta$ and apply \smartref{rule:dist_until}
            %(or \smartref{rule:dist_eventually})
            to $\beta$
            and distinguish the formula inside the \emph{next} formula in $\Bruler{\beta}$;
        \end{itemize}
        \item if $\phi$ is an $\exists$-formula ($\phi =
          \Lhid{x}{\phi'}$), then
          create a new node $l'$ as a child of $l$
          and label it as $\Tlabel{l'} = (\Tlabel{l}\setminus \{\phi\})\cup\{ \phi' \}$
    \end{itemize}
    \item If $\Tlabel{l}$ is a set of elementary formulas,
    apply the $\nextop{}$ operator: create a new node $l'$ as child of $l$ and label it as
    $\Tlabel{l'} = \nextop{\Tlabel{l}}$.
\end{enumerate}    

The construction terminates when every branch is marked.

\end{definition}

By construction, each stage in the systematic tableau $\Tname$ for
$\Phi$ is \absat.
In order to ensure termination of the algorithm, it is necessary to
use a \emph{fair} strategy to distinguish eventualities, in the sense
that every eventuality in an open branch must be distinguished at
some point. This assumption and the fact that, given a finite set
of initial formulas, there exist only a
finite set of possible labels in a systematic tableau, imply termination.

It is worth noticing that, by the application of the rules in
\smartref{fig:alpha_rules}, when both $\phi$ and $\neg\phi$ belong to the
labeling of a stage in a branch $b$, then any branch prefixed by $b$
is closed. Moreover, by construction, non-fulfilled undistinguished eventualities in a
branch are kept %at least 
until they are fulfilled or they become distinguished.

One key result of the tableau in \cite{GaintzarainHLNO09} and that we
borrow is that if a distinguished eventuality is not fulfilled in an
expanded branch $b$, then we can mark the branch as closed. This is
because if we apply \smartref{rule:dist_until}, then we get a
contradiction with the context of the eventuality.
 
As a conclusion, we have that every distinguished eventuality in a
cyclic branch $b$ of $\Tname$
    is fulfilled just since if it were unfulfilled, then $b$ would be
    closed (thus not cyclic).
Also, by construction and the above properties, $b$ is open if and only if (1) the last node of $b$ contains
only constraint formulas, or (2) $b$ is cyclic and all its
eventualities are fulfilled in $b$.

\begin{lemma}\label{lem:termination}
By using a fair strategy and
    given as input a finite set $\Phi \subseteq \csltl$,    the algorithm of \smartref{def:alg_tabl},
    terminates and builds an expanded tableau for $\Tname$.
\end{lemma}

\subsection{Soundness and completeness}\label{sec:tabl_sound_compl}

Let us now show that the proposed algorithm
is sound and complete for proving the satisfiability/unsatisfiability
of $\csltl$ formulas.

\begin{theorem}[soundness]\label{th:sound_tabl}
    If there exists a closed systematic tableau for $\Phi\subseteq\csltl$, then $\Phi$ is unsatisfiable.
\end{theorem}

In order to prove completeness, we need to define an auxiliary
function $\Tstores{}$ that, given a sequence of stages, builds a
suitable conditional trace which joins all the accumulated information
in a stage at each time instant. % To define $\Tstores$, 
We abuse of
notation and write $\ecrs$ the empty sequence of stages. Recall that
$\otimes$ is the join operation of the constraint system and
$\bigotimes\emptyset = \CStrue$.
%returns the correspondent conditional trace.
%
\begin{align*}
    &\Tstores{\ecrs} = \ecrs\\*
    &\Tstores{s \cdot S} =
    \csC{C}{\emptyset}{C} \cdot \Tstores{S} %\\*
%    &
\qquad\text{where}\ C = \bigotimes \set{c}{c\in\CSys, c\in\Tlabel{n}, n\in s}
\end{align*}  

By definition of $\nextop{}$, which in our case propagates the
constraints from one stage to the following, the conditional trace
$r$ resulting of applying $\Tstores{}$ to a sequence of stages $S$
is monotone. Furthermore, since all the negative conditions are empty,
$r$ is also consistent.

We show that, given a systematic tableau $\Tname$
built for $\Phi$, we can compute a model for $\Phi$ from every open branch $b$ in $\Tname$.

\begin{lemma}\label{lem:tabl_model}
  Let $b$ be an \emph{open expanded} branch in the systematic tableau
  $\Tname$ for $\Phi\subseteq\csltl$. Given the sequence of stages $S$
  in $\Tpath{b}$, then
  $\asat{\Tstores{S}}{\Phi}$.
\end{lemma}

\begin{theorem}[refutational completeness]\label{th:ref_compl}
  If $\Phi\subseteq\csltl$ is unsatisfiable, then there exists a
  closed tableau for $\Phi$.
\end{theorem}

\begin{theorem}[completeness]\label{th:compl_tabl}
    If $\Phi\subseteq\csltl$ is satisfiable,
    then there exists a finite open tableau for $\Phi$.
\end{theorem}

\subsection{Application of the Tableau}

We are interested in checking the validity of a formula of the form
$\Limpl{\psi}{\phi}$. Our strategy is to build the tableau for its
negation $\Tname[\Lneg{(\Limpl{\psi}{\phi})}]$ so that, if
$\Tname[\Lneg{(\Limpl{\psi}{\phi})}]$ is closed, meaning that
$\Lneg{(\Limpl{\psi}{\phi})}$ is unsatisfiable, then our implication
is valid.  

Let us show two examples of construction of the systematic tableaux
for two formulas of this kind.

\begin{example}\label{ex:validity_check}
    Consider the formula of our guiding example %(introduced in \smartref{ex:simpleDiag})
    $\Limpl{\Lhid{x}{\phi}}{\Leventually{(y=1)}}$,
    where
    $\phi = \Ldisj{(\Lconj{y=1}{\Lconj{\Lnext{x=5}}{\Lnext{(\Leventually{y=1})}}})}
    {(\Lconj{\Lneg{y=1}}{\Lnext{y=1}})}$.
    The tableau in \smartref{fig:tabl1},
    with $\Tlabel{\mathit{root}}=\Lconj{\Lhid{x}{\phi}}{\Lalways{\Lneg{(y=1)}}}$ shows its validity.
\begin{figure}[htd!]
    \centering{
    \begin{tikzpicture}[scale=0.5]
        %nodes
        \draw (0,0) node (root) {$\underline{\Lconj{\Lhid{x}{\phi}}{\Lalways{(\Lneg{y=1})}}}$};
        \draw (root)+(0.5,-1) node (alpha0) {$\alpha$};
        \draw (root)+(0,-2) node (root2) {$\underline{\Lhid{x}{\phi}}, \Lalways{(\Lneg{y=1})}$};
        \draw (root2)+(0.5,-1) node (exists) {$\exists$};
        \draw (root2)+(0,-2) node (zero) {$\underline{\phi}, \Lalways{(\Lneg{y=1})}$};
        \draw (zero)+(0,-1) node (beta) {$\beta$};
        \draw (zero)+(-5,-2) node (one) {$\underline{\Lconj{\Lconj{y=1}{\Lnext{x=5}}}{\Lnext{(\Leventually{y=1})}}}, \Lalways{(\Lneg{y=1})}$};
        \draw (one)+(0.5,-1) node (alpha1) {$\alpha$};
        \draw (one)+(0,-2) node (three) {$y=1,\Lnext{x=5}, \Lnext{(\Leventually{y=1})}, \underline{\Lalways{(\Lneg{y=1})}}$};
        \draw (three)+(0.5,-1) node (alpha5) {$\alpha$};
        \draw (three)+(0,-2.5) node (three2) {$\begin{aligned}[t]
        &y=1, \Lnext{x=5}, \Lnext{(\Leventually{y=1})},\\
        &\Lneg{y=1}, \Lnext{\Lalways{(\Lneg{y=1})}}\end{aligned}$};
        \draw (three2)+(0,-1.2) node (closedthree) {$\closedleaf$};
        \draw (zero)+(5,-2) node (two) {$\underline{\Lconj{\Lneg{y=1}}{\Lnext{y=1}}}, \Lalways{(\Lneg{y=1})}$};
        \draw (two)+(0.5,-1) node (alpha2) {$\alpha$};
        \draw (two)+(0,-2) node (five) {$\Lneg{y=1}, \Lnext{y=1}, \underline{\Lalways{(\Lneg{y=1})}}$};
        \draw (five)+(0.5,-1) node (alpha3) {$\alpha$};
        \draw (five)+(0,-2) node (six) {$\Lneg{y=1}, \Lnext{y=1}, \Lnext{(\Lalways{(\Lneg{y=1})})}$};
        \draw (six)+(0.5,-1) node (next) {$X$};
        \draw (six)+(0,-2) node (seven) {$y=1, \underline{\Lalways{(\Lneg{y=1})}}$};
        \draw (seven)+(0.5,-1) node (alpha4) {$\alpha$};
        \draw (seven)+(0,-2) node (eight) {$y=1, \Lneg{y=1}, \Lnext{(\Lalways{(\Lneg{y=1})})}$};
        \draw (eight)+(0,-0.7) node (closedeight) {$\closedleaf$};
        %arrows
        \draw[-latex] (root) -- (root2);
        \draw[-latex] (root2) -- (zero);
        \draw[-latex] (zero) -- (one);
        \draw[-latex] (zero) -- (two);
        \draw[-latex] (one) -- (three);
        \draw[-latex] (three) -- (three2);
        \draw[-latex] (two) -- (five);
        \draw[-latex] (five) -- (six);
        \draw[-latex] (six) -- (seven);%NEXT la voglio grande la freccia
        \draw[-latex] (seven) -- (eight);
    \end{tikzpicture}
    }%
    \caption{Tableau for $\Limpl{\Lhid{x}{\phi}}{\Leventually{y=1}}$
    of \smartref{ex:validity_check}.}
    \label{fig:tabl1}
\end{figure}
    Arrows labeled with $\alpha$ and $\beta$
    correspond to the application of
    $\alpha$ and $\beta$ rules, % (\smartref{fig:alpha_rules}) 
\resp{};
    arrows labeled with $X$ represent the application of the
    $\nextop{}$ operator.
    Finally, arrows labeled with $\exists$
    correspond to the elimination of the existential quantification.
%    for the formula $\Lhid{x}{\phi}$.

In the example, the first step uses the rule for the
conjunction. Then, the second step involves
the elimination of the existential quantification for $\Lhid{x}{\phi}$.
Since the formula $\Lalways{(\Lneg{y=1})}$, which
represents the context, does not contain information about $x$,
$\Lhid{x}{\phi}$ can be replaced with $\phi$.
\begin{extendedvers}(\smartref{cor:exists_corr}).\end{extendedvers}
The
formula $\phi$ is then selected for a $\beta$ step (disjunction). The branch
on the left is closed after two steps since $y=1$ and $\Lneg{y=1}$
both belong to the node labeling. 

The branch on the right, first flattens the conjunction and then
applies the next rule. Note that the negation of a constraint is
not kept in the following time instant. We recall that negation means ``not entailment''
(in contrast to meaning that \emph{the contrary is true}),
thus, in the future, the constraint could become true.

Since both branches are closed, we know that the formula
$\Lconj{\Lhid{x}{\phi}}{\Lalways{\Lneg{(y=1)}}}$
is not satisfiable, thus its negation
$\Limpl{\Lhid{x}{\phi}}{\Leventually{(y=1)}}$
is valid. 

In the context of
abstract diagnosis, this proves that the program is abstractly correct
\wrt\ the LTL specification.
\end{example} 

\begin{example}\label{ex:fail_check}
    Now, suppose that we want to check,
    for the program introduced in \smartref{ex:simple1},
    that the constraint $y=1$ is always
    entailed by the store.
    The corresponding specification is
    $\SF'(\apcall{p}{y}) = \Lalways{(y=1)}$.
    
    The \csltl-semantics $\FDd{}{}$ for $\apcall{p}{y}$ using the
    given specification as interpretation is %given by the formula
     $\Lhid{x}{\big( \Ldisj{(\Lconj{y=1}{\Lconj{\Lnext{x=5}}{\Lnext{(\Lalways{y=1})}}})}
        {(\Lconj{\Lneg{y=1}}{\Lnext{y=1}})} \big)}$.
    Let us abbreviate the body of the existential quantification as
    $\phi'$.
    %= \Ldisj{(\Lconj{y=1}{\Lconj{\Lnext{x=5}}{\Lnext{(\Lalways{y=1})}}})}
    %{(\Lconj{\Lneg{y=1}}{\Lnext{y=1}})}$.
    To check whether the
    process $\apcall{p}{y}$ is correct \wrt{} the property, %specification
    we have to show that $\Lhid{x}{\phi'} \leqF \Lalways{(y=1)}$
    is valid.
    
    \smartref{fig:tabl3} shows part of the (finite) tableau that
    proves the satisfiability of the formula
    $\Lconj{\Lhid{x}{\phi'}}{\Leventually{(\Lneg{y=1}})}$.  This means
    that its negation, $\Lhid{x}{\phi'} \leqF \Lalways{(y=1)}$, is not
    valid.  In the context of abstract diagnosis, although the formula
    is actually not satisfied by the program, because of the loss of precision
    due to the approximation, this is only a warning about the
    possible incorrectness of the program \wrt\ the LTL specification.
    
\begin{extendedvers}
    Notice that
    the second step
    involves the elimination
    of the existential quantification $\Lhid{x}{}$.
    Furthermore,
    \smartref{rule:dist_until} is applied twice to
    deal with the distinguished eventuality
    $\Leventually{(\Lneg{y=1})}$.
 \end{extendedvers}   
    \begin{figure}[htp]
        \centering{%\fbox{
        \begin{tikzpicture}[scale=0.5]
            %nodes
            \draw (0,0) node (root) {$\underline{\Lconj{\Lhid{x}{\phi'}}{\Leventually{(\Lneg{y=1})}}}$};
            \draw (root)+(0.5,-1) node (alpha0) {$\alpha$};
            \draw (root)+(0,-2) node (root2) {$\underline{\Lhid{x}{\phi}}, \Leventually{(\Lneg{y=1})}$};
            \draw (root2)+(0.5,-1) node (exists) {$\exists$};
            \draw (root2)+(0,-2) node (zero) {$\underline{\phi'},
            \Leventually{(\Lneg{y=1})}$};
            \draw (zero)+(0,-1) node (beta) {$\beta$};
            \draw (zero)+(-5,-2) node (one) {$\underline{\Lconj{\Lconj{y=1}{\Lnext{x=5}}}{\Lnext{(\Lalways{y=1})}}}, \Leventually{(\Lneg{y=1})}$};
            \draw (one)+(0.5,-1) node (alpha1) {$\alpha$};
            \draw (one)+(0,-2) node (three) {$y=1, \Lnext{x=5}, \Lnext{(\Lalways{y=1})}, \underline{\Leventually{(\Lneg{y=1})}}$};
            \draw (three)+(-0.4,-1) node (beta3) {$\beta$};
            \draw (three)+(-3,-2.2) node (four) {$\begin{aligned}
            &y=1, \Lnext{x=5},\\ &\Lnext{(\Lalways{y=1})}, \Lneg{y=1}\end{aligned}$};
            \draw (four)+(0,-1.2) node (closedfour) {$\closedleaf$};
            \draw (three)+(0.5,-5) node (five) {$\begin{aligned}
             &y=1, \Lnext{(\Lalways{y=1})},\Lnext{x=5}, \\ &\Luntil{(\Ldisj{\Lneg{y=1}}{\Lneg{\Lnext{(\Lalways{y=1})}}})}{(\Lneg{y=1})}\end{aligned}$};
            \draw (five)+(-2,-2) coordinate (five1);
            \draw (five)+(2,-2) coordinate (five2);
            \draw (zero)+(6,-2) node (two) {$\underline{\Lconj{\Lneg{y=1}}{\Lnext{y=1}}}, \Leventually{(\Lneg{y=1})}$};
            \draw (two)+(0.5,-1) node (alpha2) {$\alpha$};
            \draw (two)+(0,-2) node (six) {$\Lneg{y=1}, \Lnext{y=1}, \underline{\Leventually{(\Lneg{y=1})}}$};
            \draw (six)+(-0.2,-1) node (beta2) {$\beta$};
            \draw (six)+(-2,-2) node (seven) {$\Lneg{y=1}, \Lnext{y=1}$};
            \draw (seven)+(0.5,-1) node (sevennext) {$X$};
            \draw (seven)+(0,-2) node (eight) {$y=1$};
            \draw (eight)+(0,-0.7) node (openeight) {$\openleaf$};
            \draw (six)+(2,-5.5) node (nine) {$\begin{aligned}
            \Lneg{y=1}, \Lnext{y=1}&,\\ \Lnext{\Luntil{(\Ldisj{y=1}{\Lneg{\Lnext{y=1}}})}{&(\Lneg{y=1})}}
            \end{aligned}$};
            \draw (nine)+(-2,-2) coordinate (nine1);
            \draw (nine)+(2,-2) coordinate (nine2);
            %arrows
            \draw[-latex] (root) -- (root2);
            \draw[-latex] (root2) -- (zero);
            \draw[-latex] (zero) -- (one);
            \draw[-latex] (zero) -- (two);
            \draw[-latex] (one) -- (three);
            \draw[-latex] (three) -- (four);
            \draw[-latex] (three) -- (five);
            \draw[-latex] (two) -- (six);
            \draw[dashed] (five) -- (five1);
            \draw[dashed] (five) -- (five2);
            \draw[-latex] (six) -- (seven);
            \draw[-latex] (seven) -- (eight);
            \draw[-latex] (six) -- (nine);
            \draw[dashed] (nine) -- (nine1);
            \draw[dashed] (nine) -- (nine2);
        \end{tikzpicture}
        }%}%
        \caption{Tableau for $\Limpl{\Lhid{x}{\phi'}}{\Lalways{y=1}}$
        of \smartref{ex:fail_check}}
        \label{fig:tabl3}
    \end{figure}
\end{example}

\section{How to handle streams}

In \tccp\,
streams are used to model imperative-style variables.
In this context, when specifying an intended behavior,
we are interested in the \emph{current} value in a stream $S$ (\ie\
the last instantiated value in the stream),
denoted as $S \streq  \mathit{value}$.

In order to deal in our tableau with this special kind of constraint, 
% verify these specifications,
we define a stream simplification function
$\Tstream{}$ that transforms a formula $\phi$ %with streams
\begin{extendedvers}(that is usually generated from a program by means of $\FDd{}{}$)\end{extendedvers}
into a new formula $\phi'$
that contains information only about the last instantiated
value of the streams in $\phi$. We show that the transformation preserves satisfiability.
\begin{extendedvers}$\phi'$
and abstract away from the information
about the auxiliary
substreams used in the program.
\end{extendedvers}

The function $\Tstream{}$ uses two auxiliary functions\begin{extendedvers}: $\streamDep{}$ and $\streamHead{}{}$\end{extendedvers}.
$\streamDep{\phi}$ generates
a set of dependencies on the form $(S,S')$, where $S$ is a stream and
$S'$ its tail
in the formula $\phi$,
while
$\streamHead{S}{D}$ returns, 
given a set of dependencies $D$, the
first name given to the stream $S$.
For instance, for $\phi = \Luntil{(\Lconj{(S = [c \mid S'])}{(S'
    = [c' \mid S''])})}{(T = [d \mid T'])}$, 
$\streamDep{\phi} = \{ (S,S'), (S',S''), (T,T') \}$
and $\streamHead{S''}{\streamDep{\phi}} = S$. Formally,

\begin{equation*}\label{eq:streamDep}
    \streamDep{\phi} \dfn
    \begin{cases}
        \emptyset & \text{if}\ \phi = \Ltrue,\ \phi = \Lfalse\ \text{or}\ \phi = c \\
        (S,S') & \text{if}\ \phi = (S =[c \mid S'])\\
        \streamDep{\phi_1} & \text{if}\ \phi = \Lhid{x}{\phi_1}\ \text{or}\ \phi = \Lnext{\phi_1}\\
        \streamDep{\phi_1} \cup \streamDep{\phi_2} & \text{if}\ \phi = \Lconj{\phi_1}{\phi_2}\
        \text{or}\ \phi = \Luntil{\phi_1}{\phi_2} 
    \end{cases}
\end{equation*}

\begin{equation*}\label{eq:streamHead}
    \streamHead{S}{D} \dfn
    \begin{cases}
        S & \text{if}\ D=\emptyset \\ 
        \streamHead{S'}{D'} & \text{if}\ D= \{ (S',S) \} \cup D' \\ 
        \streamHead{S}{D'} & \text{if}\ D= \{ (T',T) \} \cup D'\ \text{and}\ T\neq S  \\ 
    \end{cases}
\end{equation*}

\begin{definition}\label{def:Tstream}
Let $\phi,\phi_1,\phi_2 \in\csltl$, $c\in\CSys$ and $S,S',R,R'$ streams of $\CSys$. $\Tstream{}$ is defined inductively as follows.
\begin{equation*}\label{eq:Tstream}
    \Tstream{\phi} \dfn
    \begin{cases}
        \phi & \text{if}\ \phi = \Ltrue,\ \phi = \Lfalse\ \text{or}\ \phi = c \\
        S\streq c & \text{if}\ \phi = (S' =[c \mid S''])\ \text{and}\
        \streamHead{S'}{\streamDep{\phi}} = S \\
        \Lneg{\Tstream{\phi_1}} & \text{if}\ \phi = \Lneg{\phi_1} \\
        \Lconj{\Tstream{\phi_1}}{\Tstream{\phi_2}} & \text{if}\
        \phi = \Lconj{\phi_1}{\phi_2}\\
        \Lhid{x}{\Tstream{\phi_1}} & \text{if}\ \phi = \Lhid{x}{\phi_1}\\
        \Lnext{\Tstream{\phi_1}} & \text{if}\ \phi = \Lnext{\phi_1}\\
        \Luntil{\Tstream{\phi_1}}{\Tstream{\phi_2}} & \text{if}\
        \phi = \Luntil{\phi_1}{\phi_2} 
    \end{cases}    
\end{equation*}
\end{definition}

\begin{example}
    Consider $\phi = (\Lconj{\Lconj{C=[\mathit{near}\mid C']}{\Lnext{(C'=[\mathit{out}\mid C''])}}}
    {G = [\mathit{down}\mid G']})$, then %by applying the stream simplification
    $\Tstream{\phi} = (\Lconj{\Lconj{C \streq \mathit{near}}{\Lnext{(C \streq  \mathit{out})}}}
    {G \streq  \mathit{down}})$.
\end{example}

The transformation
$\Tstream{}$ preserves satisfiability, thus it can be used
to preprocess the initial formula
before applying the tableau method.
%to check its validity/satisfiability.

\begin{lemma}\label{lem:Tstream}
    Let $\phi\in\csltl$, $\phi$ is satisfiable $\iff$ $\Tstream{\phi}$
    is satisfiable.
\end{lemma}    

If we need to use this transformation, then %When we handle streams 
the $\nextop{}$ operator  must be
slighly modified: % in the following way:
$\nextop{\Phi} \dfn \set{\phi}{\Lnext{\phi} \in \Phi} \cup
\set{\Lneg{\phi}}{\Lneg{\Lnext{\phi}} \in \Phi} \cup \set{c}{c\in\Phi,
c\in \CSys } \cup
\set{S \streq  c_1}{S \streq  c_1 \in\Phi\ \text{and}\ \not\exists c_2\
\text{such that}\ \Lnext{(S \streq c_2)} \in \Phi\ \text{and}\
c_1\neq c_2}$.
Intuitively, the constraints on the form $S \streq c_1$ are propagated only if
in the next time instant the tail of $S$
has not be instantiated with a different constraint $c_2$.
%It is easy to notice that 
This %extension of the 
definition of $\nextop{}$
preserves satisfiability of \smartref{lem:step_corr}.

\section{Conclusions and Future Work}\label{sec:conclusions}

In this paper, we have introduced a decision procedure for %a class of
\csltl\ formulas. \csltl\ is a linear temporal logic that replaces
propositional formulas by constraint formulas, thus in order to
determine the validity of a formula with no temporal constructs, it
uses the entailment relation of the underlying constraint system.

This decision procedure is the last step of a method to validate
\ltl{} formulas for \tccp{} programs. 
It is an adaptation of the tableau defined in
\cite{GaintzarainHLN08,GaintzarainHLNO09}. The main differences of our
algorithm \wrt\ the propositional case are
due to the constraint nature of the behavior of \tccp{}.

A Constraint Linear Temporal Logic is defined in \cite{Valencia05} for
the verification of a different timed concurrent language, called
\ntcc{}, which shares with \tccp{} the concurrent constraint nature
and the non-monotonic behavior.  A fragment of the proposed logic, the
restricted negation fragment where negation is only allowed for state
formulas, is shown to be decidable. However, no efficient decision
procedure is given. \begin{extendedvers} (apart from the proof itself)\end{extendedvers}. Moreover, the
verification results are given for the locally-independent fragment of
\ntcc{}, which avoids the non-monotonicity of the original
language. In contrast, our abstract diagnosis technique 
checks temporal properties for the full \tccp{} language.

As future work, we plan to implement this algorithm and integrate it
with the verification method.
We also plan to explore other instances of the method based on logics for
which decision procedures or (semi)automatic tools exists.

% \draft{\bibliographystyle{\draftbibstyle}}
% \official{\bibliographystyle{\finalbibstyle}}
% \bibliography{new-laura,new-alicia,biblio}%

\begin{extendedvers}
\appendix

\section{Abstract semantics evaluation function for agents}

The following function is the core definition for the correct abstract
semantics for \tccp{} in the domain of \csltl{} formulas. Actually,
this version of the semantics is an instance of the general framework
in which we are restricted to a decidable subset of the \csltl{} logic. For this
reason, the semantics for the choice agent is a correct, but not
precise, semantics of the agent's behavior. 

\begin{definition}[\csltl\ abstract Semantics]% Evaluation Function for Agents]
    \label{def:semFAa} 
    
    Given $A\in\Agents$ and $\FI \in \interpF$, we define the \emph{\csltl\
    semantics evaluation} $\FAa{A}{\FI}$ by structural induction as
    follows.
    \begin{subequations}
        \begin{align}
            %skip
            & \FAa{\askip}{\FI} \dfn \CStrue
            \label{eq:FAa_skip}
            \\
            %tell
            & \FAa{\atell{c}}{\FI} \dfn \Lnext{c}
            \label{eq:FAa_tell}
            \\
            %ask 
    &\FAa{\asumask{n}{c}{A}}{\FI} \dfn \Ldisj{\Ldisj{_{i=1}^{n}
    (\Lconj{c_i}{\Lnext{\FAa{A_i}{\FI}}})}{}} {(\Lconj{_{i=1}^{n}
    \Lneg{c_i}}{})}
    \label{eq:FAa_ask2}
            \\
            %now
            & \FAa{\anow{c}{A_1}{A_2}}{\FI} \dfn
            \Ldisj{(\Lconj{c}{\FAa{A_1}{\FI}})}{(\Lconj{\Lneg{c}}{\FAa{A_2}{\FI}})}
            \label{eq:FAa_now}
            \\
            %parallel
            & \FAa{A_1 \parallel A_2}{\FI} \dfn \Lconj{\FAa{A_1}{\FI}}{\FAa{A_2}{\FI}}
            \label{eq:FAa_par}
            \\
            %parallel
            & \FAa{\ahiding{x}{A}}{\FI} \dfn \Lhid{x}{\FAa{A}{\FI}}
            \label{eq:FAa_hid}
            \\
            %procedure call
            & \FAa{ \mgc{p}{x}{} }{\FI} \dfn \Lnext{\FI( \mgc{p}{x}{} )}
            \label{eq:FAa_pcall}
        \end{align}
    \end{subequations}
    Let $\P\in\Progs$.  We define the (monotonic) immediate consequence
    operator $\FDd{\P}{} \colon \interpF \to \interpF$ as
    \begin{equation*}
        \FDd{\P}{\FI}( \mgc{p}{x}{} ) \dfn \lubF{ \set*{ \FAa{A}{\FI} }{
        \mgrule{p}{x}{A} \in D} }{}
        %\label{eq:FD}
    \end{equation*}
\end{definition}
\end{extendedvers}
\end{document}